\documentclass[]{aa}

\newcommand{\mstar}{$M_\star$}

\newcommand{\hb}{H$\beta$}
\newcommand{\ha}{H$\alpha$}

\newcommand{\oiii}{[O\,{\footnotesize III}]}
\newcommand{\oii}{[O\,{\footnotesize II}]}
\newcommand{\nii}{[N\,{\footnotesize II}]}

\newcommand{\dn}{$D_n 4000$}
\newcommand{\name}{AURORA-LQG1}

\usepackage{graphicx}
\usepackage{txfonts}
\usepackage{lipsum}
\usepackage{subcaption}         
\usepackage{lscape}             
\usepackage{placeins}           
\usepackage{booktabs}
\usepackage[dvipsnames]{xcolor}
\usepackage[colorlinks=true, citecolor=blue]{hyperref}
                   
\begin{document}
\title{Dynamical properties and star formation history of a low-mass quenched galaxy at Cosmic Noon}
\author{K.~Ito\inst{1,2}
\and F.~Valentino\inst{1,2}
\and W.~M.~Baker\inst{3}
\and G.~Brammer\inst{1,4}
\and R.~Gottumukkala\inst{1,4}
\and T.~Kakimoto\inst{5,6}
\and C.~D.~P.~Lagos \inst{7, 8, 1}
\and M.~Onodera\inst{5,9}
\and A.~Pensabene\inst{1,2}
\and G.~Scarpe\inst{1,2}
\and M.~Tanaka\inst{5,6}
\and K.~E.~Whitaker\inst{10,1}
\and N.~A.~Reddy\inst{11}
\and R.~L.~Sanders\inst{12}
\and A.~E.~Shapley\inst{13}
}

\institute{Cosmic Dawn Center (DAWN), Copenhagen, Denmark
\and 
DTU Space, Technical University of Denmark, Elektrovej 327, DK2800 Kgs. Lyngby, Denmark
\and
DARK, Niels Bohr Institute, University of Copenhagen, Jagtvej 155A, DK-2200 Copenhagen, Denmark 
\and
Niels Bohr Institute, University of Copenhagen, Jagtvej 128, 2200 Copenhagen N, Denmark
\and
Department of Astronomical Science, The Graduate University for Advanced Studies, SOKENDAI, 2-21-1 Osawa, Mitaka, Tokyo 181-8588, Japan
\and 
National Astronomical Observatory of Japan, 2-21-1 Osawa, Mitaka, Tokyo 181-8588, Japan
\and
International Centre for Radio Astronomy Research (ICRAR), M468, University of Western Australia, 35 Stirling Hwy, Crawley, WA 6009, Australia
\and
ARC Centre of Excellence for All Sky Astrophysics in 3 Dimensions (ASTRO 3D)
\and
Subaru Telescope, National Astronomical Observatory of Japan, National Institutes of Natural Sciences (NINS), 650 North A'ohoku Place, Hilo, HI 96720, USA
\and
Department of Astronomy, University of Massachusetts, Amherst, MA 01003, USA
\and
Department of Physics \& Astronomy, University of California, Riverside, 900 University Avenue, Riverside, CA 92521, USA
\and 
Department of Physics and Astronomy, University of Kentucky, 505 Rose Street, Lexington, KY 40506, USA
\and 
Department of Physics \& Astronomy, University of California, Los Angeles, 430 Portola Plaza, Los Angeles, CA 90095, USA
}
   
 \date{Received -, Accepted -} 
  \abstract{We present the spectroscopic confirmation and in-depth analysis of \name, a low-mass quiescent galaxy at $z_{\rm spec}=2.0834$ with $\log(M_\star/M_\odot)=9.6$, observed with medium-resolution JWST/NIRSpec spectroscopy. Its stellar mass places \name\ $\sim10\times$ below the knee of the stellar mass function for quiescent galaxies at $z\sim2$. The deep medium-resolution spectrum enables the measurement of its stellar velocity dispersion ($\sigma_\star = 95_{-33}^{+38}\,{\rm km\,s^{-1}}$), the smallest value recorded among spectroscopically confirmed quiescent galaxies at $z\sim2$. Coupled with a compact size ($0.41\pm0.03\, {\rm kpc}$ in the rest-frame optical), the stellar velocity dispersion yields a dynamical mass estimate of $\log(M_{\rm dyn}/M_\odot)=9.75_{-0.38}^{+0.29}$, consistent with the stellar mass, confirming the true low-mass nature of this galaxy, and placing a first constraint on its initial mass function. Joint spectro-photometric SED fitting reveals a star formation history in which half the stellar mass was in place $\sim1\,{\rm Gyr}$ before the observed epoch, with quenching occurring $\sim0.2\,{\rm Gyr}$ prior to $z=2.08$.
  These results confirm that \name\ is genuinely quenched, rather than in a temporary phase of suppressed star formation rate.  \name\ is consistent with the mass fundamental plane at $z\sim2$, which was previously constrained only by massive quiescent systems with $M_\star\geq10^{11}\,M_\odot$ at cosmic noon. Compared with more massive counterparts at the same epoch observed with similar NIRSpec grating spectroscopy, the time since quenching for \name\ is among the shortest observed. The star formation history is broadly consistent with predictions for quiescent galaxies of similar mass and redshift in the IllustrisTNG simulation. The galaxy resides in a possible dense group-scale ($\sim50$ kpc) environment containing one companion with tentative spectroscopic redshift and five low-mass companion candidates with similar redshifts, and it is embedded in a large known protocluster on Mpc scales. A potential environmental influence on its evolution could explain the outside-in quenching suggested by the positive gradient of size with wavelength. This study demonstrates that deep JWST/NIRSpec spectroscopy enables low-mass quiescent galaxies at Cosmic Noon to be characterized with a level of detail long reserved for massive systems, offering valuable new insights into how quenching operates in these underexplored low-mass systems.}

   \maketitle

\section{Introduction}
Mass and environment are well known to correlate with galaxy quenching in the local Universe, reflecting a wide range of physical mechanisms and timescales involved \citep{ypeng_2010}. Internal “ejective’’ processes -- such as gas outflows driven by active galactic nuclei (AGNs) or supernovae (SNe) -- can shut down star formation rapidly on $<0.1-0.5$ Gyr timescales \citep[e.g.,][]{Croton_2006, Fabian_2012}, whereas “preventive’’ mechanisms like morphological stabilization act more slowly ($\gtrsim1$ Gyr). Low-mass galaxies are additionally susceptible to external processes: ram-pressure stripping can quench them on similarly short timescales ($<0.1-0.5$ Gyr, \citealp{gunn_1972}), while starvation operates more gradually over $\sim1-5$ Gyr in dense environments \citep[][]{peng_2015,cortese_2021}.
\begin{figure*}
    \centering
    \includegraphics[width=1\linewidth]{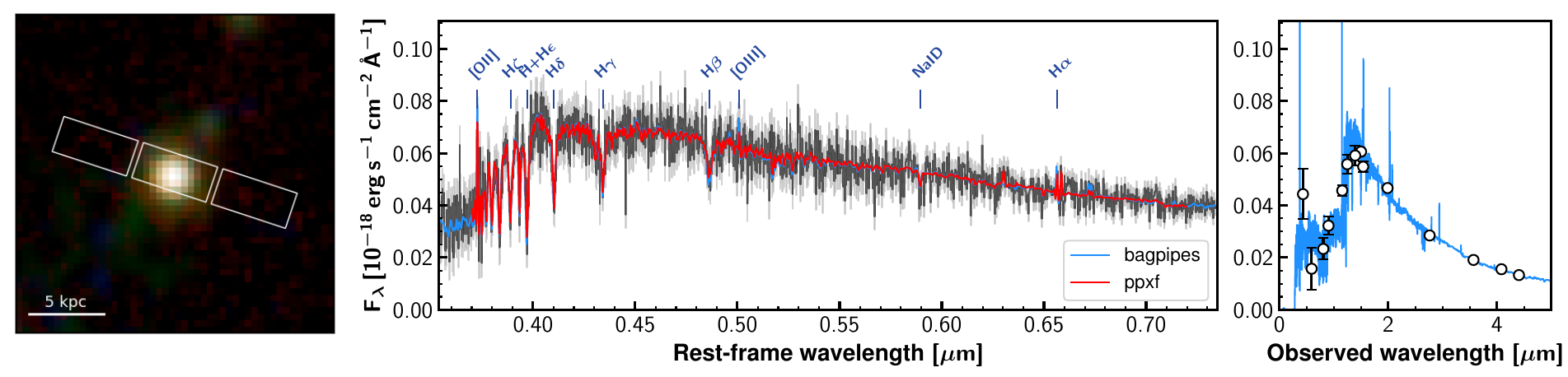}
    \caption{Left panel: $2'' \times 2''$ JWST/NIRCam composite image of \name. Images with the F115W, F150W, and F444W filters are used as the blue, green, and red channels. The white boxes indicate the NIRSpec/MSA slitlets positioning. Middle panel: JWST/NIRSpec spectrum of \name. The black line and gray shaded region correspond to the observed spectrum and its $1\sigma$ uncertainty, respectively. The red and blue lines correspond to the best-fit from {\sc pPXF} (Section \ref{subsec:ppxf}) and {\sc Bagpipes} (Section \ref{subsec:stellarpop}), respectively. Right panel: Photometric SED of \name. The blue line is the best-fit SED obtained with {\sc Bagpipes}. }
    \label{fig:spec}
\end{figure*}

Much of our current knowledge of how star formation ceases in galaxies at high redshift comes from massive objects, whose quenching is thought to be dominated by internal mechanisms and can be probed with high-quality near-infrared (NIR) spectroscopy. Over the past decade, spectroscopic observations have confirmed a substantial population of red, quiescent galaxies at early times. This effort has been transformed by JWST: its exceptional sensitivity enables clear detections of stellar continua in massive ($\log{(M_\star/M_\odot)}>10$) quiescent galaxies at $z\sim2-7$, with absorption features and overall spectral shapes yielding robust measurements of their star formation histories, stellar velocity dispersions, chemical compositions, and surrounding environment \citep[e.g.,][]{glazebrook_2017, schreiber_2018c, tanaka_2019, valentino_2020a, forrest_2020b, esdaile_2021, forrest_2022, marchesini_2023, kakimoto_2024,belli_2024,davies_2024,wu_2024,antwi-danso_2025, beverage_2024, carnall_2024, glazebrook_2024, Slob_2024, nanayakkara_2025, deGraaff_2025, weibel_2025, valentino_2025, baker_2024, Baker_2025, McConachie_2025, Ito_2025c, Slob_2025}.

Photometric studies, especially those using deep JWST imaging, have also established the presence of a population of quiescent galaxies extending to much lower stellar masses, revealing their abundance and morphological properties \citep[][]{weaver_2023, cutler_2024, Alberts_2024, Baker_2025b, Shuntov_2025, Hamadouche_2025}. The resulting stellar mass functions (SMFs) rise from high masses to a peak around $M_\star\sim10^{10.5}\,M_\odot$ at $z\sim2$, while at lower masses the mechanisms traditionally associated with the onset of environmental effects progressively build up the power-law tail of the classical Schechter function. Stellar masses even 1 dex below the peak of the SMF of quiescent galaxies at $z\sim2$ hold the promise of constraining such environmental effects and are becoming key to calibrating simulations and models \citep{lagos_2025,Baker_2025b,Shuntov_2025}. Nevertheless, this mass range is almost unexplored spectroscopically. Moreover, the rare spectroscopic confirmations of quiescent galaxies with $\log{(M_\star/M_\odot)}<10$ at $z\gtrsim2$ rely on low-resolution spectroscopy (NIRSpec/PRISM or grism observations; \citealt{Sandles_2023,Sato_2024,baker_2024,Baker_2025}), largely because of the long integration times required. This severely limits the physical analysis of the formation and evolution of these systems compared with the level of detail attained for their more massive counterparts, and it completely prevents a first assessment of their dynamical properties.

This paper reports the characterization of a low-mass quiescent galaxy, \name, with $\log{(M_\star/M_\odot)}=9.6$ at $z=2.08$, about $10\times$ less massive than the peak of the SMF of quiescent galaxies at this redshift \citep[e.g.,][]{weaver_2023}. The analysis is based on deep JWST/NIRSpec medium-resolution spectroscopy combined with JWST/NIRCam and HST imaging. The medium spectral resolution is crucial for confirming the low-mass nature of this system through a first estimate of its dynamical mass and a robust spectrophotometric modeling of its star formation history, achieving a level of quality comparable to that obtained for more massive counterparts. At lower redshifts, less massive galaxies are known to have formed more recently than their massive counterparts, a phenomenon referred to as  ``downsizing'' \citep[][]{thomas_2005, Thomas_2010, gallazzi_2005, Gallazzi_2025}. Observations and analyses of low-mass quiescent galaxies as the one presented here therefore provide an ideal opportunity to begin exploring the mass dependence of star-formation histories at Cosmic Noon, in light of results for massive systems in the literature \citep[e.g.,][]{park_2024, baker_2024, nanayakkara_2025}.

This work is structured as follows. Section \ref{sec:data} summarizes the data set used in this paper. In Section \ref{sec:spectralanalysis}, we present the spectral analysis, deriving the kinematical properties (Section \ref{subsec:ppxf}) and the stellar population properties (Section \ref{subsec:stellarpop}). Section \ref{sec:dynamical} compares the dynamical properties with well-established scaling relations. In Section \ref{sec:discussion}, we discuss the dependence of the star-formation history on stellar mass and the group-like environment surrounding our target. Throughout this work, we use the AB magnitude system \citep{oke_1983}. We adopt a $\Lambda$CDM cosmology with $\Omega_{\rm m} = 0.3$, $\Omega_{\Lambda} = 0.7$, and $H_0 = 70\,\mathrm{km\,s^{-1}\,Mpc^{-1}}$. 


\section{Target selection and data}
\label{sec:data}
We selected our target galaxy, \name\ (Figure \ref{fig:spec}), as part of the DeepDive program and archival search for quiescent galaxies with NIRSpec/Micro-shatter array (MSA) medium or high-resolution spectroscopy at high redshift \citep{Ito_2025c}. \name\ 
is in the COSMOS field \citep{scoville_2007} and it was selected based on its high \dn\ value ($D_n4000 =1.35\pm0.03$, defined as in  \citealt{balogh_1999}), its $UVJ$ rest-frame colors ($U-V=1.18\pm0.09$, $V-J=0.63\pm0.06$), falling within the selection box in \citet{williams_2009} with 0.2 dex margin, and the low specific star formation rate (sSFR) from photometric spectral energy distribution (SED) fitting, placing the galaxy more than one dex below the star formation main sequence in \cite{schreiber_2015}. We briefly summarize the data available for \name\ and the reduction here below, and we refer the reader to \cite{Ito_2025c} and references therein for all details.

We retrieve JWST/NIRCam and HST imaging data (v7) from the Dawn JWST Archive (DJA)\footnote{\url{https://dawn-cph.github.io/dja/}}. JWST/NIRCam data were obtained as part of the Public Release IMaging for Extragalactic Research (PRIMER) Survey (PID \#1837; PI: Dunlop) and reduced with \textsc{Grizli} \citep{brammer_2023, valentino_2023}. Here we use the photometry measured in circular apertures with a diameter of $0\farcs5$ and corrected to “total” values within Kron apertures \citep{kron_1980}, which were computed on the detection image obtained by combining filters in the long-wavelength channels. This study uses imaging obtained with the F090W, F115W, F150W, F200W, F277W, F356W, F410M, and F444W filters from JWST/NIRCam, as well as the F814W, F125W, F140W, and F160W bands from HST/ACS and WFC3.

JWST/NIRSpec spectroscopic data of \name\ were collected as a part of the Assembly of Ultradeep Rest-optical Observations Revealing Astrophysics program\footnote{The ID of this galaxy in the AURORA survey is 4631.} \citep[AURORA, PID \#1914, Co-PIs: A. Shapley and R. Sanders,][]{Shapley_2025}. Spectra were collected with the G140M/ F100LP, G235M/F170LP, and G395M/F290LP combinations, with exposure times of 44,204s, 28,886s, and 15,056s, respectively. Here we focus on the G140M/F100LP and G235M/F170LP spectra mapping the rest-frame optical wavelengths at $z=2$. The data were reduced using the {\sc Msaexp} software \citep{brammer_2023_msaexp, heintz_2025, degraaff_2024_rubies}. The latest version of the reduction available on DJA (v4\footnote{\url{https://doi.org/10.5281/zenodo.15472354}}, \citealt{valentino_2025}) extends the wavelength coverage of each spectrum. Once optimally extracted \citep{horne_1986}, we calibrated the spectra against flux losses and contamination from higher-order spectra with a custom correction for quiescent galaxies \citep{Ito_2025c}. The flux loss calibration was applied by anchoring the observed spectra to the total photometry with a second order polynomial function. The spectra and SED of \name\ are shown in Figure \ref{fig:spec}.

\section{Spectral analysis}
\label{sec:spectralanalysis}
\subsection{Redshift and stellar velocity dispersion}
\label{subsec:ppxf}
We conducted the stellar template fitting using the Penalized Pixel fitting algorithm (\textsc{pPXF}, \citealt{cappellari_2017,cappellari_2023}) and derived the redshift and stellar velocity dispersion ($\sigma_\star$) of \name. We adopted the Flexible Stellar Population Synthesis (FSPS) models \citep{conroy_2010} with MILES stellar libraries \citep{Sanchez-Blazquez_2006, falcon-barroso_2011} as templates. We allowed the stellar ages to be younger than the age of the Universe at the redshift of the target and the metallicity to range between $\mathrm{[M/H]}=-2$ and $0.5$. 

A careful treatment of the NIRSpec instrumental resolution is essential for an accurate stellar velocity dispersion measurement. Recent work has shown that the in-flight spectral resolution is higher than the pre-launch estimates \citep[][]{deGraaff_2024_ISM}. Here we adopt the JDox resolution curve\footnote{\url{https://jwst-docs.stsci.edu/jwst-near-infrared-spectrograph/nirspec-instrumentation/nirspec-dispersers-and-filters\#gsc.tab=0}} scaled up by a factor of 1.3 \citep{deGraaff_2025,valentino_2025}. This factor is consistent with the prescription of \citet{Slob_2024} for sources of size comparable to \name\ (Section \ref{subsec:morph}). Although the exact correction also depends mildly on source position, the resulting variation is at the $\sim10$\% level and it does not affect our analysis. The wavelength-dependent resolution curve is incorporated into {\sc pPXF}, and the templates are convolved accordingly. 

Following previous studies \citep[][]{Scholz-Diaz_2022}, we derived the kinematical and stellar population properties using a two-step process. Both steps were performed over the wavelength range $0.37\,{\rm \mu m}<\lambda< 0.72\,{\rm \mu m}$ due to the resolution of the MILES templates. In the first step, we constrained the stellar kinematics by estimating the velocity offset (i.e., a refined redshift) and the stellar velocity dispersion. To mitigate potential mismatches between the observed spectrum and the stellar templates, we flattened both by modeling the spectral shape with a third-order B-spline function and dividing the spectra by it prior to fitting. During this step, we modeled only the stellar continuum and masked strong emission lines (\ha, \nii, \oiii, \hb, and \oii), as well as the sodium absorption doublet (NaI D). The latter may be possibly affected by absorption from the neutral interstellar gas and outflows, commonly observed at $z\sim2$ \citep[][]{davies_2024}. A second-order additive polynomial correction was also included. This step was iterated 1000 times by adding random noise to the observed spectra to estimate the uncertainties on $z$ and $\sigma_\star$. In the second step, we reran {\sc ppxf}, simultaneously modeling the stellar continuum and emission lines while fixing the kinematics to the values obtained in the first step. A second-order multiplicative polynomial correction was included this time, and the NaI~D absorption was again masked. This second step was also iterated 1000 times to estimate the uncertainties on the stellar population properties, such as stellar age and metallicity, as well as on emission line fluxes.

Figure \ref{fig:spec} shows the best-fit model from {\sc pPXF}. The best-fit redshift is $z_{\rm spec} = 2.0834_{-0.0002}^{+0.0002}$, used hereafter. We estimate a stellar velocity dispersion of $\sigma_\star = 95_{-33}^{+38}\, {\rm km\, s^{-1}}$, which is the lowest stellar velocity dispersion reported so far at $z\geq2$ \citep[e.g.,][]{belli_2014,belli_2017a,tanaka_2019,Stockmann_2020,esdaile_2021,forrest_2022,Slob_2025}. The measured emission line fluxes are also summarized in Table \ref{tab:summary}. The \ha\ line is detected at a significance of $7\sigma$, allowing us to estimate the instantaneous star formation rate following \citet{kennicutt-evans_2012}, assuming a negligible contribution from AGN. This assumption is supported by the \oiii/\hb\ and \nii/\ha\ ratios, which, despite their large uncertainties ($\log{\rm ([OIII]\lambda5007/H\beta)}=-0.22_{-0.29}^{+0.21}$ and $\log{\rm ([NII]\lambda6583/H\alpha)}=-0.26_{-0.13}^{+0.11}$, respectively), suggesting no significant AGN contribution according to the BPT diagnostic diagram \citep{kewley_2013l}. Given the large uncertainties on the H$\beta$ line flux and the negligible dust extinction on the stellar continuum inferred from the SED fitting (see Section \ref{subsec:stellarpop}), we do apply a dust-attenuation correction to the line fluxes. The resulting instantaneous star formation rate is ${\rm SFR}_{\rm H\alpha}= 0.08_{-0.01}^{+0.01} M_\odot\, {\rm yr^{-1}}$. 

\subsection{SED modeling}
\label{subsec:stellarpop}
We used {\sc Bagpipes} \citep{carnall_2018,Carnall_2019b} to model the spectra at $0.35\, {\rm \mu m}<\lambda< 0.74\,{\rm \mu m}$ and photometry simultaneously. We adopted the stellar population models from \citet{bruzual_2003}, with the MILES stellar spectral library and stellar evolutionary tracks from \citet{bressan_2012} assuming the initial mass function (IMF) from \citet{kroupa_2001}; the dust attenuation law from \citet{calzetti_2000}, and the grid of nebular emission lines with a constant ionization parameter $\log{U}=-3$. We assumed two different star formation histories: a double power-law function \citep{carnall_2019} and a non-parametric continuity model \citep{leja_2019_sfh}. In addition to the star formation history, the stellar metallicity and velocity dispersion are set as free parameters. The redshift was fixed to the spectroscopic estimate from {\sc pPXF}. We also included a 2nd-order polynomial function in the fit to correct the mismatch between the stellar templates and spectra and an extra white noise scaling factor. The effect of the varying resolution of JWST grating spectra was accounted for by inputting the resolution curve described above. Consistent with our approach to model the spectrum with {\sc pPXF}, we masked out the spectrum around the \ha, \nii, \oiii, \hb, and \oii\ emission lines and the NaI D absorption feature. Priors and the allowed range of free parameters are summarized in Appendix \ref{sec:config_Bapgipes}. 

We report the best-fit parameters in Table \ref{tab:summary}. The best-fit models of the spectra and the photometry are shown in Figure \ref{fig:spec}. With a double power-law parameterization of the SFH, we obtain a stellar mass of $\log{(M_\star/M_\odot)} = 9.612_{-0.012}^{+0.009}$ and a SFR averaged over 100 Myr of ${\rm SFR}=0.14_{-0.03}^{+0.03}\, M_\odot\, {\rm yr^{-1}}$. The error bars on \mstar\ reflect the statistical uncertainties in the fitting process, while systematic uncertainties of $\sim0.2$ dex are known to affect the estimate of this parameter \citep{pacifici_2023}. Although the time scale of interest is different, the SFR from the SED modeling is broadly consistent with the instantaneous SFR from H$\alpha$ (${\rm SFR}_{\rm H\alpha}= 0.08_{-0.01}^{+0.01} M_\odot\, {\rm yr^{-1}}$, see Section \ref{subsec:ppxf}). This corresponds to an sSFR being 1 dex lower than for main sequence galaxies with the same stellar mass at $z=2$ \citep[e.g.,][]{schreiber_2015,popesso_2023}, and lower than $0.2/t_{\rm age}$, where $t_{\rm age}$ is the age of the Universe at that redshift. This makes \name\ one of the least massive quiescent galaxies at $z\geq2$ observed with NIRSpec grating spectra \citep[for those observed with PRISM spectra, see][]{baker_2024,Sandles_2023,Sato_2024,Baker_2025}. We obtain consistent results with the non-parametric SFH (Table \ref{tab:summary}). The stellar velocity dispersion derived from {\sc Bagpipes} is consistent with that from {\sc pPXF}.

\subsection{Star formation history and metal enrichment}
\begin{figure}
    \centering
    \includegraphics[width=1\linewidth]{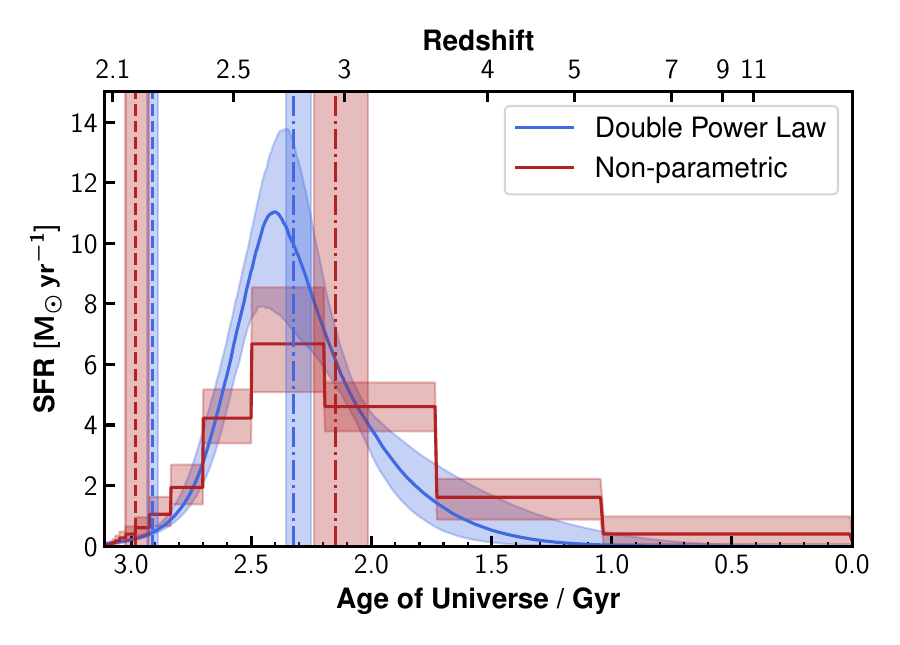}
    \caption{Star formation history of \name. The solid lines and shaded regions in blue and red colors indicate the medians and 16-84\% percentile ranges obtained by modeling the spectra and photometry with {\sc Bagpipes}, and assuming a double power-law and a non-parametric model, respectively. The dashed-dotted and dashed vertical lines for each color correspond to the formation and quenching times for each SFH model.}
    \label{fig:SFH}
\end{figure}

Figure \ref{fig:SFH} shows the SFH of \name\ derived from the spectro-photometric modeling with \textsc{Bagpipes}. The double power-law and the non-parametric models show overall similar shapes. The time of the formation, i.e., the time after the Big Bang when half of the observed stellar mass was formed, are $t_{\rm form}= 2.33_{-0.06}^{+0.03}$ Gyr and $t_{\rm form}= 2.14_{-0.13}^{+0.09}$ Gyr for the double power-law and the non-parametric model, respectively. This time of the formation means that the half of the \name was formed $\sim 1\,{\rm Gyr}$ before the observed redshift, which is also suggested from the stellar age estimate from the {\sc pPXF} fitting ($1.55_{-0.34}^{+0.31}$ Gyr). We also computed the time of the quenching, defined as the epoch when $\mathrm{sSFR} \leq 0.2/t_{\rm age}$ after the formation time \citep[e.g.,][]{park_2024,baker_2024}. Also in this case, we find that both SFHs parameterizations return similar quenching time ($2.91_{-0.02}^{+0.02}$\,Gyr and $2.98_{-0.06}^{+0.04}$\,Gyr after the Big Bang for the double power-law and the non-parametric forms, respectively). The shape of the spectra and the ensuing formation and quenching timescales are thus similar to those of more massive quiescent galaxies \citep[][]{park_2024,Slob_2024,nanayakkara_2024,carnall_2023b,carnall_2024}, and markedly different from those of lower-mass, bluer galaxies likely being in a transitory phase of suppressed star formation (``mini-quenched" systems, \citealt{strait_2023, looser_2024, baker_2025_miniqg}). In Section \ref{subsec:SFHsim}, we will compare in more detail the modeled SFHs with those of other massive quiescent galaxies at similar redshifts and in a cosmological simulation.

The modeling with \textsc{Bagpipes} yields stellar metallicities of ${\rm [M/H]}=0.06_{-0.06}^{+0.04}\,$ and ${\rm [M/H]}=-0.07_{-0.06}^{+0.04}$ for the double power-law and the non-parametric SFH models, respectively. These values are $\sim0.3-0.4$ dex higher than the metallicity derived with {\sc pPXF} (${\rm [M/H]} = -0.41_{-0.25}^{+0.22}$). This discrepancy may arise from the different assumptions about the star formation history \citep{deGraaff_2025} or from differences in the datasets used in the modeling, as both photometry and spectroscopy were included when running {\sc Bagpipes}, whereas only the spectrum was used with {\sc pPXF}. The limited sensitivity to weak metal absorption features is imprinted on the large uncertainty of the metallicity measurements, such that all three estimates are broadly consistent with the stellar metallicities of quiescent galaxies at similar stellar mass in the local Universe \citep[${\rm [M/H]} \sim -0.5$ with a 1$\sigma$ scatter of $0.5$ dex;][]{gallazzi_2005,Mattolini_2025,Gallazzi_2025}. This suggests that \name\ was already significantly metal-enriched by $z\sim2$, reaching levels comparable to those observed in local quiescent galaxies, in agreement with studies of massive quiescent systems at similar redshift \citep[][]{beverage_2025}.
\begin{table}
    \caption{Summary of physical properties of \name.}\label{tab:summary}
    \centering
    \begin{tabular}{cc}
        \hline\hline
        Parameter & Value\\
        \hline
        R.A. & 150.156402\\
        Decl. & 2.224525\\
        \hline
        \multicolumn{2}{c}{From pPXF\tablefootmark{a}}\\  
        \hline
        $z$ & $2.0834_{-0.0002}^{+0.0002}$  \\
        $\sigma_\star$ [${\rm km\, s^{-1}}$] & $95_{-33}^{+38}$ \\
        ${\rm [M/H]}$ & $-0.41_{-0.25}^{+0.22}$ \\
        $t_{\rm age}$ [Gyr] & $1.55_{-0.34}^{+0.31}$ \\
        $f_{\rm H\beta}\ {\rm [erg\, s^{-1}\, cm^{-2}]}$ & $2.4_{-0.7}^{+0.7}\times10^{-19}$ \\
        $f_{\rm H\alpha}\ {\rm [erg\, s^{-1}\, cm^{-2}]}$ & $5.0_{-0.7}^{+0.7}\times10^{-19}$ \\
        $f_{\rm [OII]3726}\ {\rm [erg\, s^{-1}\, cm^{-2}]}$ & $4.3_{-1.0}^{+0.9}\times10^{-19}$ \\
        $f_{\rm [OII]3729}\ {\rm [erg\, s^{-1}\, cm^{-2}]}$ & $1.9_{-0.9}^{+0.9}\times10^{-19}$ \\
        $f_{\rm [OIII]}\ {\rm [erg\, s^{-1}\, cm^{-2}]}$\tablefootmark{b} & $2.2_{-0.9}^{+0.9}\times10^{-19}$ \\
        $f_{\rm [OI]}\ {\rm [erg\, s^{-1}\, cm^{-2}]}$ & $2.1_{-1.0}^{+1.0}\times10^{-19}$ \\
        $f_{\rm [NII]}\ {\rm [erg\, s^{-1}\, cm^{-2}]}$\tablefootmark{b} & $3.5_{-0.8}^{+0.8}\times10^{-19}$ \\
        ${\rm SFR}_{\rm H\alpha}$ [$M_\odot\, {\rm yr^{-1}}$]& $0.08_{-0.01}^{+0.01}$ \\
        \hline
        \multicolumn{2}{c}{From Bagpipes with DPL SFH}\\
        \hline
        $\log{(M_\star/M_\odot)}$ & $9.612_{-0.009}^{+0.012}$\\
        SFR [$M_\odot\, {\rm yr^{-1}}$] & $0.14_{-0.03}^{+0.03}$\\
        sSFR [${\rm yr^{-1}}$] & $3.4_{-0.6}^{+0.7}\times10^{-11}$\\
        $A_V$ [mag]& $0.02_{-0.01}^{+0.03}$\\
        ${\rm [M/H]}$ & $0.06_{-0.06}^{+0.04}$ \\
        $t_{\rm form}$\tablefootmark{c} [Gyr] & $2.33_{-0.06}^{+0.03}$ \\
        $t_{\rm quench}$\tablefootmark{c} [Gyr] & $2.91_{-0.02}^{+0.02}$\\
        \hline
        \multicolumn{2}{c}{From Bagpipes with Non-parametric SFH}\\
        \hline
        $\log{(M_\star/M_\odot)}$ & $9.630_{-0.010}^{+0.013}$\\
        SFR [$M_\odot\, {\rm yr^{-1}}$] & $0.24_{-0.10}^{+0.11}$\\
        sSFR [${\rm yr^{-1}}$] & $5.7_{-2.5}^{+2.6}\times10^{-11}$\\
        $A_V$ [mag]& $0.03_{-0.02}^{+0.03}$\\
        ${\rm [M/H]}$ & $-0.07_{-0.06}^{+0.04}$ \\
        $t_{\rm form}$\tablefootmark{c} [Gyr] & $2.14_{-0.13}^{+0.09}$ \\
        $t_{\rm quench}$\tablefootmark{c} [Gyr] & $2.98_{-0.06}^{+0.04}$\\
        \hline
        \multicolumn{2}{c}{Morphological fitting in F150W}\\
        \hline
        $R_e\tablefootmark{d}$ [kpc] & $0.41_{-0.03}^{+0.03}$\\
        $n$ & $3.04_{-0.45}^{+0.52}$\\
        $q$\tablefootmark{e} & $0.85_{-0.06}^{+0.08}$\\
        \hline
        $\log{(M_{\rm dyn}/M_\odot)}$ & $9.75_{-0.38}^{+0.29}$\\
        \hline\hline
    \end{tabular}
    \tablefoot{
    \tablefoottext{a}{Only fluxes of emission lines detected more than $2\sigma$ are shown.}
    \tablefoottext{b}{Total flux of both lines in the doublet.}
    \tablefoottext{c}{Time since the Big Bang.}
    \tablefoottext{d}{Half light radius along the semi-major axis.}
    \tablefoottext{e}{Ratio between the semi-minor and the semi-major axes.}

}
\end{table}

\section{Dynamical properties}
\label{sec:dynamical}
\subsection{Morphologies, sizes, and their gradients}
\label{subsec:morph}
We modeled the surface brightness of \name\ in the available JWST/NIRCam images taken with the F115W, F150W, F200W, F277W, F356W, and F444W filters with a S\'{e}rsic profile \citep{sersic_1963}. We used the Bayesian code {\sc PySersic} \citep{Pasha2023}\footnote{\url{https://github.com/pysersic/pysersic}}. We inputted 
$5$\arcsec $\times 5$\arcsec ($42\times 42$ kpc) images centered around \name\ and the point spread functions constructed for the DJA mosaics presented in \cite{Genin2025}. The fitting was conducted individually and separately for each filter. The images, models, and residuals are shown in Appendix \ref{app:morph}.

Figure \ref{fig:morph} shows the derived half-light semi-major radius and S\'ersic index as a function of wavelengths. The size at rest-frame $0.5\,{\rm \mu m}$, covered by the F150W filter, is $0.41_{-0.03}^{+0.03}\ {\rm kpc}$. This value is significantly smaller than the average sizes of quiescent galaxies at the same stellar mass and redshift \citep[e.g., $\sim1.4$\, kpc according to the size-mass relation at $1.75<z<2.25$,][]{Hamadouche_2025}, those of more massive quiescent galaxies \citep[$\sim 0.9$\, kpc for quiescent galaxies at $z\sim2.5$ with $\log{(M_\star/M_\odot)}=10.5$ on median,][]{Yang_2025} and star-forming galaxies with similar stellar mass \citep[$1.5$\, kpc,][]{vanderwel_2014, Yang_2025}. Considering the line ratio derived in Section \ref{subsec:ppxf}, this compact size is not likely due to the significant contribution from a point-like AGN. Moreover, the sizes have a positive correlation with wavelength. Fitting a power-law to the size as a function of the rest-frame wavelength yields a positive gradient with a slope of $\gamma=\Delta\log{r_{\rm eff}}/\Delta \log{\lambda_{\rm rest}}=0.12\pm0.05$, which is in contrast with average trends recorded for more massive quiescent galaxies at the similar redshift, showing a negative correlation \citep[$\gamma \sim-0.3$,][]{vanderwel_2014,ito_2024}. At $z<1$, the strength of size gradients correlates with stellar mass, with more pronounced positive size–wavelength trends observed at lower masses \citep{Kawinwanichakij_2021}. The result for \name\ is in line with these expectations, but now at $z\sim2$. This positive correlation implies that older stellar populations, traced by longer wavelengths, are more spatially extended than the younger one, assuming a negligible impact from dust attenuation and metallicity. We estimate a S\'ersic index of $n\sim3$ across all filters, indicating that \name\ has a bulge-like spheroidal morphology. This value agrees with the median S\'ersic index of quiescent galaxies with $\log{(M_\star/M_\odot)}>10$ at $z\sim2-3$ \citep[e.g.,][]{vanderWel_2012,Martorano_2025}. We will return to the morphology of \name\ in Section \ref{subsec:env}.

\begin{figure}
    \centering
    \includegraphics[width=0.8\linewidth]{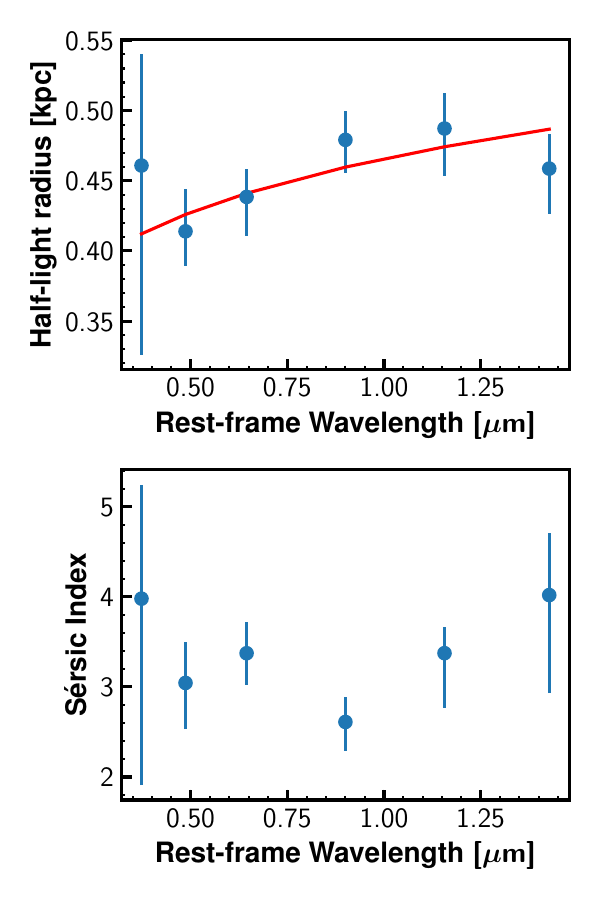}
    \caption{Upper panel: Half-light radius as a function of the rest-frame wavelength. The red line corresponds to the best-fit power-law model. Bottom panel: S\'ersic index as a function of the rest-frame wavelength. }
    \label{fig:morph}
\end{figure}

\subsection{Dynamical mass and constraints on the IMF}
\label{subsec:mdyn}
Combining the stellar velocity dispersion and morphological properties, we derived the dynamical mass of \name\ as follows:
 \begin{equation}
     M_{\rm dyn} = \frac{\beta(n)\sigma_\star^2r_{\rm eff}}{G},
     \label{eq:mdyn}
 \end{equation}
where $n$, $\sigma_\star$, $r_{\rm eff}$, and $G$ are the S\'ersic index, the stellar velocity dispersion, the effective radius, and the gravitational constant, respectively. Here we adopt the observed stellar velocity dispersion for $\sigma_\star$, but we note that the measured value may include a non-negligible, yet unconstrained, contribution from rotational motion, as many massive quiescent galaxies at high redshift are reported to exhibit significant rotation \citep[][]{Toft_2017, d'eugenio_2023, Slob_2025}. We employ the effective radius and S\'ersic index measured in the F150W image, mapping the rest-frame optical wavelength, for consistency with the literature \citep[][]{mendel_2020,Stockmann_2020,forrest_2022,Slob_2025}. The parameter $\beta$ is a function of the S\'{e}rsic index defined as $\beta (n) = 8.87-0.831\,n+0.0241\,n^2$ \citep{cappellari_2006}. We note that the dynamical mass does not change significantly, even if we consider the homology correction based on the axis ratio proposed in \citet{vanderwel_2022}, due to a high axis ratio of \name\ ($q=0.85_{-0.06}^{+0.08}$).

\begin{figure}
    \centering
    \includegraphics[width=1\linewidth]{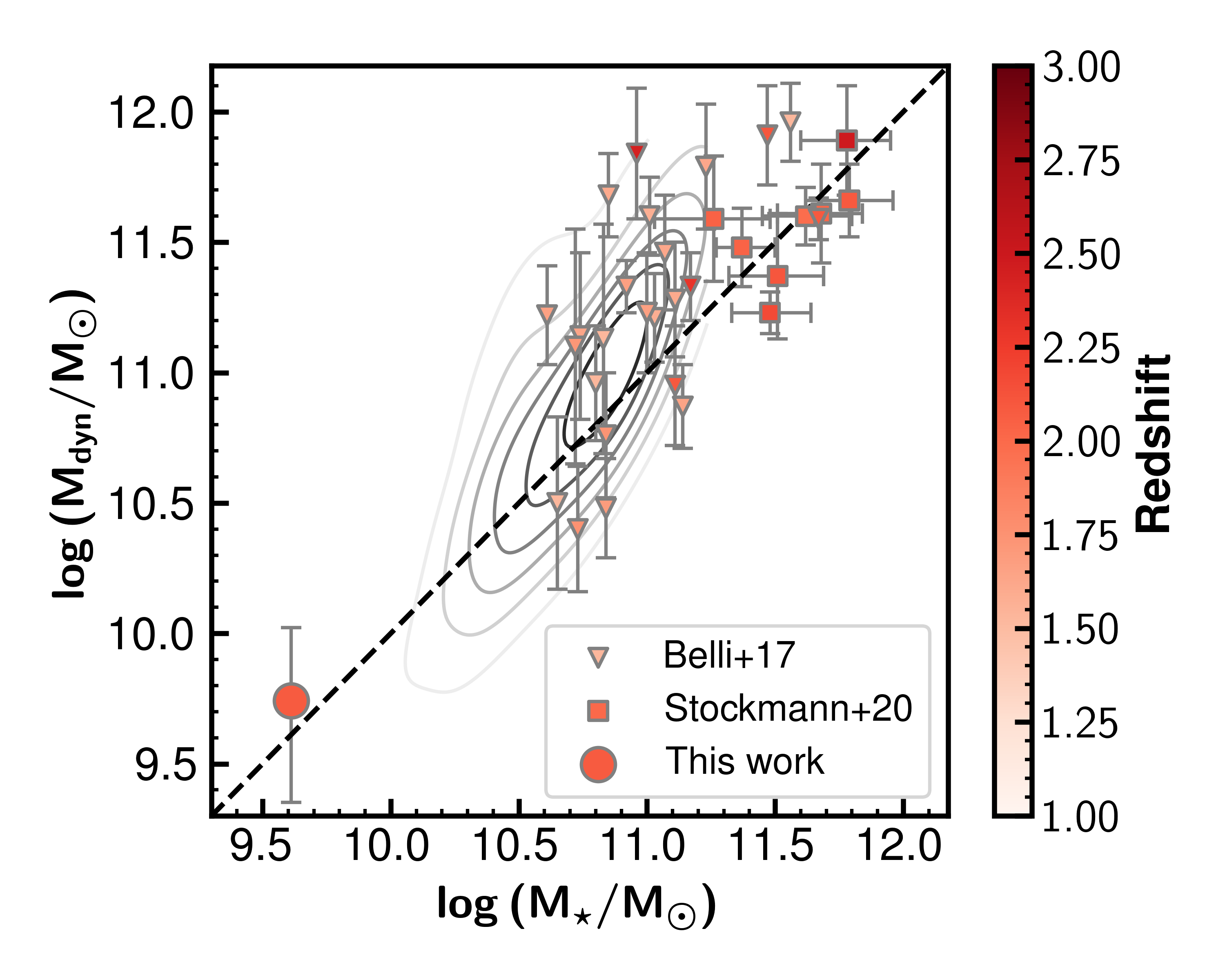}
    \caption{Relation between stellar mass and dynamical mass for quiescent galaxies at $z=2$. The large filled circle marks the location of \name. Dynamical mass estimates for quiescent galaxies at $1<z<3$ based on ground-based observations available in the literature are shown for comparison. Triangles and squares indicate the samples at $1.5<z_{\rm spec}<2.4$ from \citet{belli_2017a} and at $2<z_{\rm spec}<2.7$ from \citet{Stockmann_2020}, respectively. The symbols are color-coded according to the redshift. The contours correspond to the distribution of dynamical mass and stellar mass of quiescent galaxies observed with JWST/NIRSpec and compiled in \citet{Ito_2025c}, measured in the same manner as \name\ (Ito et al. in prep.). The black dashed line corresponds to the one-to-one relation.}
    \label{fig:Mdyn}
\end{figure}

We estimate a dynamical mass of $\log{(M_{\rm dyn}/M_\odot)} = 9.75_{-0.38}^{+0.29}$. This estimate is among the lowest values reported for quiescent galaxies at $z\geq1$ (Figure \ref{fig:Mdyn}) and it confirms the true low-mass nature of \name. The comparison with the stellar mass estimate also allows us to place a first constraint on the IMF for quiescent galaxies in this stellar mass, under the geometrical assumptions in Eq. \ref{eq:mdyn}. It is because the stellar mass depends on the assumption of the IMF, and the dynamical mass can not be smaller than the stellar mass. The dynamical-to-stellar mass ratio, assuming a Kroupa IMF (Section \ref{subsec:stellarpop}), is $\log{(M_{\rm dyn}/M_\star)}=0.14_{-0.38}^{+0.29}$. Despite the large uncertainties, this value suggests that a more bottom-heavy, \cite{salpeter_1955}-like IMF, yielding a $0.3$ dex higher stellar mass, is disfavored.

\subsection{Mass Fundamental Plane}
\label{subsec:MassFP}
\begin{figure}
    \centering
    \includegraphics[width=0.9\linewidth]{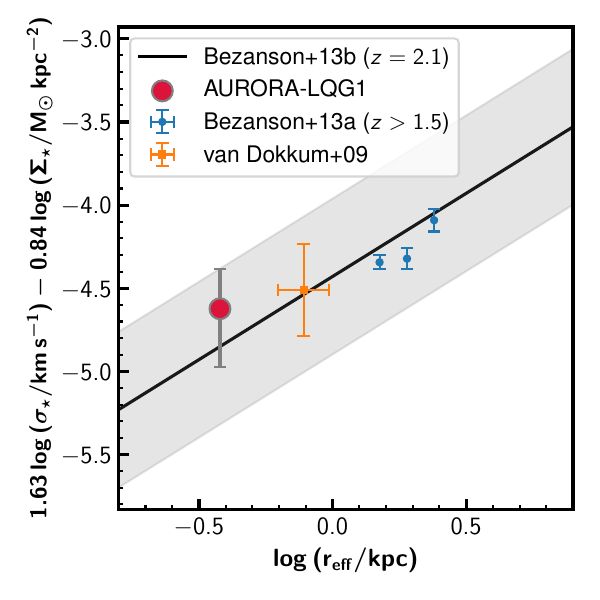}
    \caption{Mass Fundamental Plane. The location of \name\ is marked with the red filled circle, while data points for the quiescent galaxies at $z>1.5$ in \citet{Bezanson_2013} and \citet{vandokkum_2009} are shown as blue circles and an orange square, respectively. The MFP at $z=2$ from \citet{Bezanson_2013b} is shown as a black line with a gray shaded region to indicate its $1\sigma$ uncertainty.}
    \label{fig:massFP}
\end{figure}

In the local Universe, early-type galaxies follow a tight relation between their surface brightness, stellar velocity dispersion, and size, called the ``Fundamental Plane'' \citep{djorgovski_1987}. Swapping the surface brightness with the stellar mass surface brightness, the ``Mass'' Fundamental Plane (MFP) seem to be in place already at cosmic noon for massive quiescent galaxies ($\log{(M_\star/M_\odot)}\sim11$; \citealt{Bezanson_2013b}). The MFP is independent of the evolution of stellar populations \citep{Bezanson_2013b}, revealing the relationship between structural and morphological properties, and it has a negligible redshift evolution. Armed with the physical properties derived in the previous sections, we can test whether \name\ falls on the MFP. For consistency with the original definition of the MFP, we computed the circularized effective radius in F150W ($r_{\rm eff}\sqrt{q}$, where $q=0.85^{+0.08}_{-0.06}$ is the projected axis ratio from the surface brightness modeling; Table \ref{tab:summary}). Figure \ref{fig:massFP} shows that \name\ is consistent with an extension to lower masses of the MFP at $z=2$. This might hint at the fact that also low-mass quiescent galaxies ($\log{(M_\star/M_\odot)}<10$) follow the same relation between dynamical properties and morphological properties of massive galaxies at at $z=0-2$. While archival NIRSpec spectra of massive quiescent galaxies will readily test the robustness MFP in the present epoch of JWST (\citealt{Ito_2025c}, Ito et al. in prep.), lower-mass counterparts benefiting from deep medium resolution spectroscopy similar to \name\ are almost completely absent from public archives. It will thus be crucial to assemble statistical samples of $\log{(M_\star/M_\odot)}<10$ quiescent galaxies at cosmic noon to test our findings.

\section{Discussion}
\label{sec:discussion}
\subsection{Formation history and downsizing in observations and simulations}
\label{subsec:SFHsim}
At $z<1$, stellar ages are known to correlate with stellar masses such that more massive systems are systematically older (i.e., they formed earlier) than less massive ones, an effect known as ``downsizing'' \citep[e.g.,][]{thomas_2005, Thomas_2010,gallazzi_2005,Hamadouche_2023, Gallazzi_2025}. Here we test for a first possible hint of such an effect at $z\sim2$ based on the SFH of \name\ and the archival sample of massive quiescent galaxies constrained by NIRSpec grating spectra in \citet{Ito_2025c}.
For a fair comparison, we limit the redshift range to $1.9<z<2.3$ and the selection to quiescent galaxies that satisfy both the $UVJ$ color selection and the sSFR selection as \name\ does. To constrain their SFHs, we model spectra and photometry with the same setup applied to \name\ (Section \ref{sec:spectralanalysis}). After removing three sources where the modeling cannot constrain reliable quenching epochs, we retain with a sample of 15 quiescent galaxies with $M_\star>10^{10}\,M_\odot$. Their best-fit spectra and photometry are shown in Appendix \ref{app:bagpipies_massive}. 

Figure \ref{fig:SFH_tform} shows the formation and quenching times of \name\ and the archival massive quiescent galaxies. To account for the slight redshift differences in the sample, we consider the time elapsed since the formation and quenching epoch ($t_{\rm age}-t_{\rm form},\, t_{\rm age}-t_{\rm quench}$, where $t_{\rm age}$ is the age of the Universe at the observed redshift). The formation epoch of \name\ is similar to those of the massive sample. Overall, we do not observe a significant trend of downsizing in the current sample; the relatively high p-values from Spearman rank correlation test ($\rho=0.43$, $p=0.1$) does not suggest a statistically significant correlation between \mstar\ and $t_{\rm age}-t_{\rm form}$ across the $2$ dex range of stellar masses shown in Figure \ref{fig:SFH_tform}. In contrast, \name\ appears to be one of the most recently quenched galaxies in the sample at $z\sim2$, making it more akin to "post-starburst" systems than the other massive quiescent galaxies. However, we do not find a statistically significant correlation between \mstar\ and $t_{\rm age}-t_{\rm quench}$, with a Spearman rank correlation coefficient of $\rho = 0.46$ and a $p$-value of 0.07.

Next, we compare the SFH of \name\ with those of simulated quiescent galaxies. As an example, here we use the Illustris TNG-100 simulation \citep{nelson_2019}. Quiescent galaxies in TNG-100 are selected as galaxies with $\mathrm{sSFR} \leq 0.2/t_{\rm age}$. We note that all of the observed massive quiescent galaxies in the literature compilation satisfy this criterion. The star formation rate used here is the average over 100 Myr, identical to the timescale adopted in the SED modeling with \textsc{Bagpipes}. Out of 500 subhalos hosting quiescent galaxies with $\log{(M_\star/M_\odot)>9}$ at $z=2.1$ (Snapshot ID=32), we find 38 galaxies the mass range of \name\, ($9.5<\log{(M_\star/M_\odot)<10}$). We then derive the SFH of each simulated galaxy from the distribution of the formation times of the stellar particles belonging to each subhalo at $z=2.1$, sampled in bins of 10 Myr .  
The overall shape of the SFH of \name\ is in good agreement (within the $1\sigma$ uncertainty) with the median SFH of simulated systems at $9.5<\log{(M_\star/M_\odot)<10}$.  
Moreover, we find that the formation and quenching epochs of \name\ are consistent with the distributions of those of simulated quiescent galaxies in TNG-100 with similar masses and redshifts (Figure \ref{fig:SFH_tform}). On the one hand, TNG-100 predicts a downsizing effect at $z=2$, showing a weak but significant positive correlation between the stellar mass and the time since the formation epoch ($\rho=0.25$ with $p<0.01$ from a Spearman rank correlation test), i.e., more massive quiescent galaxies formed earlier. On the other hand, it does not predict a significant correlation ($\rho=-0.08$ with $p=0.06$ from a Spearman rank correlation test) between the stellar mass and the time since the quenching epoch. 

In summary, we find that the current JWST/NIRSpec grating spectroscopy sample of quiescent galaxies does not show a statistically significant downsizing signature at $z\sim2$ (i.e., correlation between the stellar mass and formation time). The Illustris TNG-100 simulation predicts the star formation history of low-mass quiescent galaxies similar to that of \name, but also a significant correlation between stellar mass and the formation time. Neither observation nor simulation predicts the significant correlation between the stellar mass and quenching time. We note that such trends vary among different simulations, as different simulations predict different star formation histories for high redshift quiescent galaxies \citep[][]{lagos_2025}. These discussions demonstrate that mass dependency of star formation history can be used to examine galaxy quenching in simulations. Again, it is critical to assemble a statistical sample of $\log{(M_\star/M_\odot)}<10$ quiescent galaxies to robustly test different models for galaxy quenching.

\begin{figure*}
    \centering
    \includegraphics[width=0.7\linewidth]{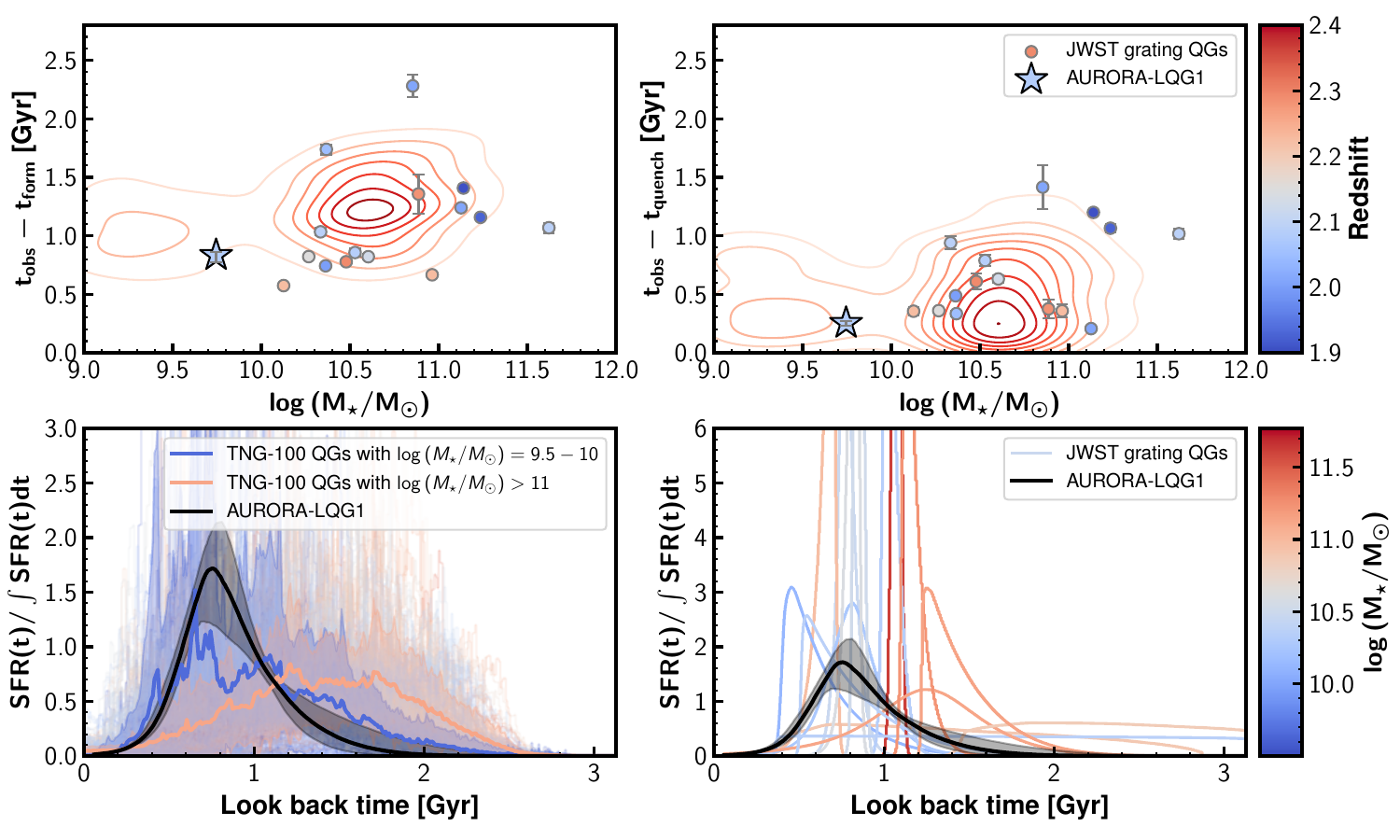}
    \caption{Top row: Look back formation ($t_{\rm age}-t_{\rm form}$, left panel) and quenching time ($t_{\rm age}-t_{\rm quench}$, right panel) as a function of stellar mass. In both panels, \name\ is marked with a star. The locations of 15 more massive quiescent galaxies at $z\sim2$ with JWST/NIRSpec grating spectra from \citet{Ito_2025c} are shown as colored circles. All symbols are color-coded according to their redshifts. The red contours indicate the distribution of quiescent galaxies in the Illustris TNG-100 simulation at $z=2.1$.
    Bottom row: Normalized star formation histories of simulated quiescent galaxies in TNG-100 at $z=2.1$ (left panel) and those of galaxies observed with NIRSpec grating spectroscopy compiled in \citet{Ito_2025c}. In the left panel, the median normalized SFHs of quiescent galaxies in TNG-100 with $9.5<\log{(M_\star/M_\odot)}<10$ and $\log{(M_\star/M_\odot)}>11$ and their 16-84\% percentiles are shown as blue and pink thick lines and shaded regions. In the right panel, the SFHs of the 15 quiescent galaxies used at upper panels is color-coded according to their stellar masses. }
    \label{fig:SFH_tform}
\end{figure*}

\subsection{The connection with the surrounding overdense environment}
\label{subsec:env}
\name\ is likely associated with a known rich proto-cluster at $z=2.095$ extended over $\sim3.7\times5$ pMpc$^2$ \citep{spitler_2012, Yuan_2014}. However, \name\ does not fall at the spatial peak of the density map reported in the literature, and its redshift is slightly lower ($\Delta z \sim 0.01$, corresponding to $\sim1000\, {\rm km\, s^{-1}}$) than the central redshift peak \citep{Yuan_2014}. 

To examine its local surrounding environment in detail, we investigate the presence of a possible localized overdensity. We first check galaxies with photometric redshift estimates consistent with \name's spectroscopic redshift ($z_{\rm spec, LQG}$) in the PRIMER-COSMOS catalog available on the DJA, based on {\sc eazy-py} \citep{brammer_2008} modeling of the photometry described in Section \ref{sec:data} \citep{valentino_2023}. We impose cuts on the stellar mass at $\log{(M_\star/M_\odot)}>8$, which is the stellar mass limit of the PRIMER-COSMOS catalog \citep[][]{valentino_2023}, and on the redshift solutions as $|z_{\rm phot,84} -  z_{\rm phot,16}|/z_{\rm phot,50}<1$ to remove contaminants and poor fits, where $z_{\rm phot, 16}$, $z_{\rm phot, 50}$, $z_{\rm phot, 84}$ are the 16, 50, 84 percentile of the probability distribution of the photometric redshift. We then searched for galaxies with photometric redshifts $|z_{\rm phot}-z_{\rm spec, LQG}|<0.1$, accounting for $1\sigma$ uncertainties on $z_{\rm phot}$. We find six galaxies within 50 pkpc from \name\ that satisfy this selection. All of them are star-forming galaxies and their stellar masses span a range between $\log{(M_\star/M_\odot)}=8.2$ and $9.5$, lower than our fiducial estimate for \name. A search in the Keck/MOSFIRE archive\footnote{\url{https://grizli-cutout.herokuapp.com/mosfire?mode=table}} available as part of the ancillary data supporting the DJA yields a tentative spectroscopic redshift for 1/6 source with $\log{(M_\star/M_\odot)}\sim9.5$ located $\sim20$\, pkpc from \name. Its MOSFIRE K-band spectrum, reduced as described in \citet{valentino_2022}, shows an emission line at $\lambda\sim2.027\, {\rm \mu m}$ with the signal-to-noise ratio $S/N=6$ (Figure \ref{fig:KeckHalpha} in Appendix \ref{app:env}). Considering this galaxy's photometric redshift at $z\sim2$, we conclude that the emission line is the H$\alpha$ line, obtaining the spectroscopic redshift to $z_{\rm spec} = 2.0878\pm0.0002$. 

To quantify the significance of the local overdensity around \name, we computed the average number of galaxies within 50 pkpc around photometrically selected galaxies with similar stellar mass ($\log{(M_\star/M_\odot)}=9.5-10$) at the same redshift ($1.98<z<2.18$). Surrounding galaxies with $\log{(M_\star/M_\odot)}>8$ are selected around each central object as for \name. For 345 galaxies with $\log{(M_\star/M_\odot)}=9.5-10$ in the PRIMER-COSMOS catalog, we find an average of $2.1\pm0.1$ galaxies within 50 pkpc. The presence of six companions around \name\ indicates that it resides in a region three times as dense as typical galaxies of similar mass. These results suggest that \name\ is not only part of a dense protocluster-scale structure but also a possibly locally overdense environment, although spectroscopic confirmation is required to robustly establish the group membership.

Recent studies on massive \citep{kubo_2021, mcconachie_2022, ito_2023, kakimoto_2024, ito_2025b, McConachie_2025, sillassen_2025} and low-mass quiescent galaxies \citep{Sandles_2023, Baker_2025} at $z\geq2$ also report that they reside in dense environments. Although it is difficult to conclude that such a dense environment is directly linked to the quenching of \name, this is in line with a possible connection between quenching of low-mass quiescent galaxies and the environment, in principle. This possibility is further supported by the high fraction of star-forming galaxies ($96\%$), as classified by the $UVJ$ color selection, in the sample used to measure the average number of surrounding galaxies.

Considering that our low-mass target \name\ was relatively rapidly quenched ($<1$ Gyr), environmental processes such as tidal or ram-pressure stripping could potentially explain its properties. However, ram-pressure stripping typically occurs in galaxy clusters with a hot intracluster medium, and since diffuse X-ray emission is not detected in this protocluster region \citep{spitler_2012}, it is less likely to be significant here. Furthermore, the positive size gradient with wavelength (Section \ref{subsec:morph}) indicates that the older stellar populations are more spatially extended than the younger stars, suggesting an ``outside-in'' quenching scenario. This contrasts with massive quiescent galaxies, which are more commonly quenched ``inside-out'' \citep[e.g.,][]{Cheng_2025_SUSPENSE}. We note that another low-mass quiescent galaxy at $z=2$ in a dense environment \citep{Sandles_2023} exhibits a similar positive color gradient. Such outside-in quenching may be driven by the environmental effects described above. Additionally, dense environments foster frequent major and minor mergers, which could increase the size of \name\ at later epochs without significantly lowering its metallicity or altering its position on the mass fundamental plane. 
\begin{figure}
    \centering
    \includegraphics[width=1\linewidth]{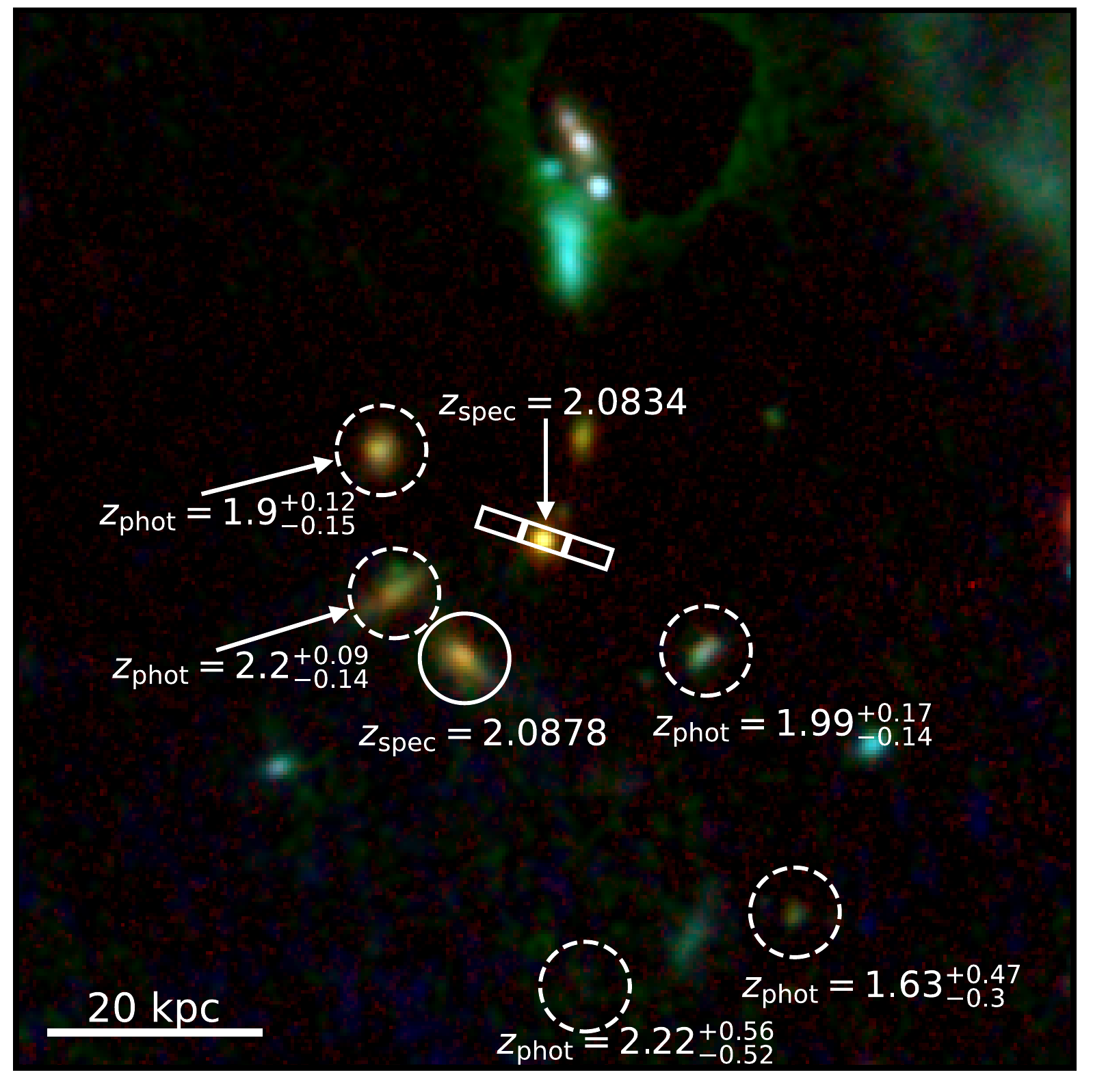}
    \caption{$12'' \times 12''$ JWST/NIRCam composite image of \name. Galaxies with spectroscopic (photometric) redshift estimates consistent with that of \name\ within their uncertainty are highlighted by solid (dashed) line circles. Their redshifts are also shown.}
    \label{fig:env}
\end{figure}
\section{Summary}
\label{sec:summary}
In this paper, we report the spectroscopic confirmation of \name, a low-mass quiescent galaxy at $z_{\rm spec} =2.0834$ with $\log{(M_\star/M_\odot)}=9.6$. This galaxy lies $10\times$ below the knee of the stellar mass function of quiescent galaxies at $z=2$, thus being one of the least massive quiescent galaxies at high redshift observed with JWST/NIRSpec grating spectroscopy so far. Analyzing the deep medium resolution spectra and photometry, we reach the following conclusions:
\begin{itemize}
    \item Its stellar velocity dispersion is measured to be $95_{-33}^{+38}\, {\rm km\, s^{-1}}$ from stellar template fitting. This value is lower than any other measurements reported for quiescent galaxies at $z\sim2$, consistent with its low stellar mass. 
    \item The spectro-photometric modeling returns a star formation history indicating that \name\ formed half of its stellar mass $\sim 1\, {\rm Gyr}$ before the time of observation, followed by quenching $\sim 0.2\, {\rm Gyr}$ prior to $z=2.08$. This makes \name\ a bona fide quenched galaxy akin to more massive counterparts in the literature, unlike bluer systems in a temporary quiescent phase.
    \item The multi-wavelength morphological analysis indicates that \name\ is a compact ($r_{\rm eff}=0.41^{+0.03}_{-0.03}$ kpc in F150W) spheroid with S\'{e}rsic index $n=3$ with a positive size gradient with wavelength, consistent with quenching occurring outside-in. 
    \item The combination of the low stellar velocity dispersion and compact size yields a dynamical mass of $\log{(M_{\rm dyn}/M_\odot)}=9.75_{-0.38}^{+0.29}$, which confirms the true low-mass nature of \name.
    \item Under standard geometrical assumptions, the dynamical-to-stellar mass ratio ($\log{(M_{\rm dyn}/M_\star)}=0.14_{-0.25}^{+0.40}$) shows that a bottom-heavy \cite{salpeter_1955} IMF is less preferable to \cite{kroupa_2001} IMF.
    \item \name\ is consistent with the extrapolation of the Mass Fundamental Plane drawn from massive quiescent galaxies with $M_\star\geq10^{11}\,M_\odot$ at $z\sim2$ \citep{Bezanson_2013b}. This suggests a possible extension of the equilibrium regulating the structure and dynamics of massive quiescent galaxies at cosmic noon down to $10-100\times$ lower masses.
    \item Comparing the derived star formation history with those of massive quiescent galaxies, \name\ is among the most recently quenched systems at $z\sim2$ observed with NIRSpec grating spectroscopy. However, we do not find a significant signature of downsizing, i.e., no statistically significant correlation between formation time and stellar mass. The star-formation history of this galaxy is broadly consistent with that of quiescent galaxies of similar mass and redshift in the Illustris TNG-100 simulation, which itself exhibits a significant correlation between stellar mass and formation time.
    \item \name\ is surrounded by a possible dense group-scale ($\sim 50$ kpc) environment, where there are five star-forming galaxies with $\log{(M_\star/M_\odot)}=8.2-9.5$ and consistent photometric redshifts, one tentatively spectroscopically confirmed. The galaxy is also part of a known larger protocluster-scale overdensity. This overdense environment may be linked to the properties and quenching of \name, particularly the possible outside-in quenching scenario, as supported by the positive size gradient with wavelength.
\end{itemize}

This study shows that deep JWST/NIRSpec medium-resolution spectroscopy of low-mass quiescent galaxies at $z\geq2$ provides access to a unique window on their physics, allowing for unambiguous confirmation of their low mass nature via dynamical analysis, for reconstructing robust star formation histories, and for placing a first constraint on critical quantities such as stellar metallicities. The assembly of statistical samples is well within the reach of JWST/NIRSpec, as our study demonstrates, and it is the only way forward to obtain typical properties of quiescent galaxies as a function of mass and environment and to constrain the representative quenching mechanism in this low-stellar mass range.


\begin{acknowledgements}
KI, FV, and PZ acknowledge support from the Independent Research Fund Denmark (DFF) under grant 3120-00043B. WMB would like to acknowledge support from DARK via the DARK Fellowship. This study was supported by JSPS KAKENHI Grant Numbers JP23K13141 and JP25K07361 and a research grant (VIL54489) from VILLUM FONDEN. This work is based on observations made with the NASA/ESA/CSA James Webb Space Telescope. The data were obtained from the Mikulski Archive for Space Telescopes at the Space Telescope Science Institute, which is operated by the Association of Universities for Research in Astronomy, Inc., under NASA contract NAS 5-03127 for JWST. These observations are associated with program \#1914. Some of the data products presented herein were retrieved from the Dawn JWST Archive (DJA). DJA is an initiative of the Cosmic Dawn Center, which is funded by the Danish National Research Foundation under grant DNRF140. 
\end{acknowledgements}

%

\bibliographystyle{aa} 
\bibliography{bib_deepdive} 

\begin{appendix}
\section{The configuration of the spectral-photometric modeling with {\sc Bagpipes}}\label{sec:config_Bapgipes}
Table \ref{tab:priors} summarizes the free parameters and priors of the spectral photometric modeling with {\sc Bagpipes} (Section~\ref{subsec:stellarpop}).
\begin{table}[]
    \centering
    \caption{Free parameters and priors for the spectro-photometric modeling with {\sc Bagpipes} .}
    \begin{tabular}{lcc}
    \toprule
    \toprule
        Free parameter & Prior & Limits \\
    \midrule
        $\sigma_\star$ & Uniform & (10, 500)\\
        $\mathrm{log}(M_{\rm formed}/M_\odot)$ & Uniform & (1, 13)\\
        $A_{\rm V}/{\rm mag}$ &                            Uniform & (0,  4)\\
        $Z / Z_\odot$ &                          Logarithmic & (0.00355,  3.55)\\
        $a$\tablefootmark{a} & Logarithmic & (0.1,10)\\
        $P_0$\tablefootmark{b} & Gaussian ($\mu=1$, $\sigma=0.1$) & (0.75,1.25)\\
        $P_1$\tablefootmark{b} & Gaussian ($\mu=0$, $\sigma=0.1$) & (-0.25,0.25)\\
        $P_2$\tablefootmark{b} & Gaussian ($\mu=0$, $\sigma=0.1$) & (-0.25,0.25)\\
        \hline
        \multicolumn{3}{c}{Parameters in the Double Power Law SFH}\\
        $\tau / \mathrm{Gyr}$ &                  Uniform & (0.1,  $t(z_{\rm obs})$)\tablefootmark{c}\\
        $\alpha$, $\beta$ &                  Logarithmic & (10$^{-2}$, 10$^{3}$)\\
        \hline
        \multicolumn{3}{c}{Parameters in the Non-parametric SFH}\\
        ${\rm dsfr}_i$ (i=1,2,...,14) &                  Student-T  & (-10, 10)\\

    \bottomrule
    \end{tabular}
    \tablefoot{
    \tablefoottext{a}{White noise scaling factor.}
    \tablefoottext{b}{Zero, first, and second-order of polynomial correction function.}
    \tablefoottext{c}{$t(z_{\rm obs})$ is the time since the Big Bang at the redshift of the sources.}

    }
    \label{tab:priors}
\end{table}

\section{Surface brightness fitting results}\label{app:morph}
\begin{figure}
    \centering
    \includegraphics[width=0.9\linewidth]{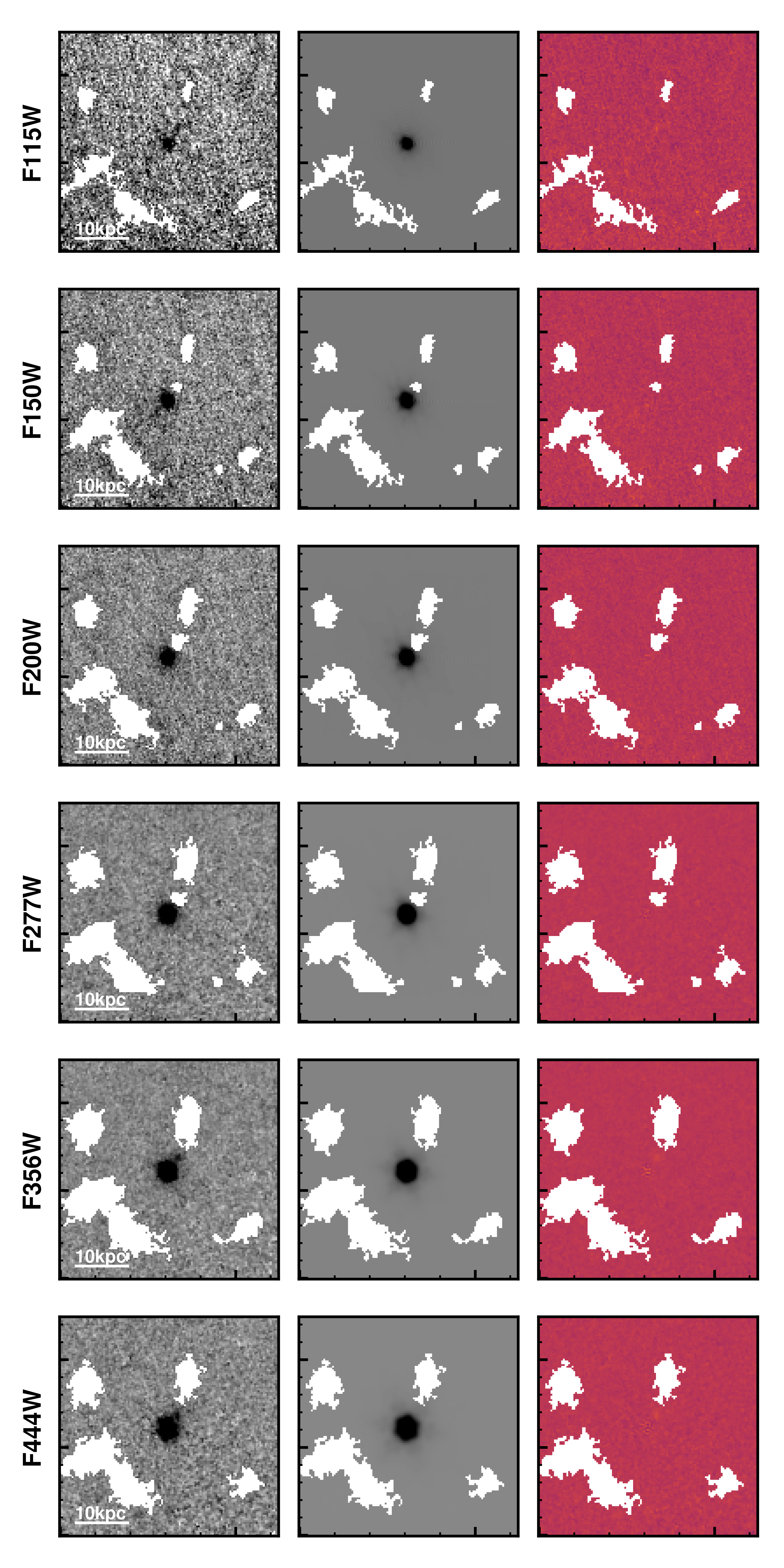}
    \caption{The surface brightness fitting results in F115W, F150W, F200W, F277W, F356W, and F444W images ($5.0\arcsec\times5.0\arcsec$). The observed images, model, and residuals are shown from left to right in each row. The white regions in each panel indicate masked nearby sources.}
    \label{fig:morphfit}
\end{figure}

Figure \ref{fig:morphfit} shows the images, S\'{e}rsic models, and residuals in all NIRCam filters presented in Section \ref{subsec:morph}.

\section{Spectro-photometric SED fitting for massive quiescent galaxies}\label{app:bagpipies_massive}
Figure \ref{fig:bagpipes_massiveQG1} and Figure \ref{fig:bagpipes_massiveQG2} show the spectra and photometry of massive quiescent galaxies used for comparison in terms of star formation history in Section \ref{subsec:SFHsim}. In these figures, the best-fit models from {\sc Bagpipes}, which give us the estimates of their star formation histories, are also shown.
\begin{figure*}
    \centering
    \includegraphics[width=0.9\linewidth]{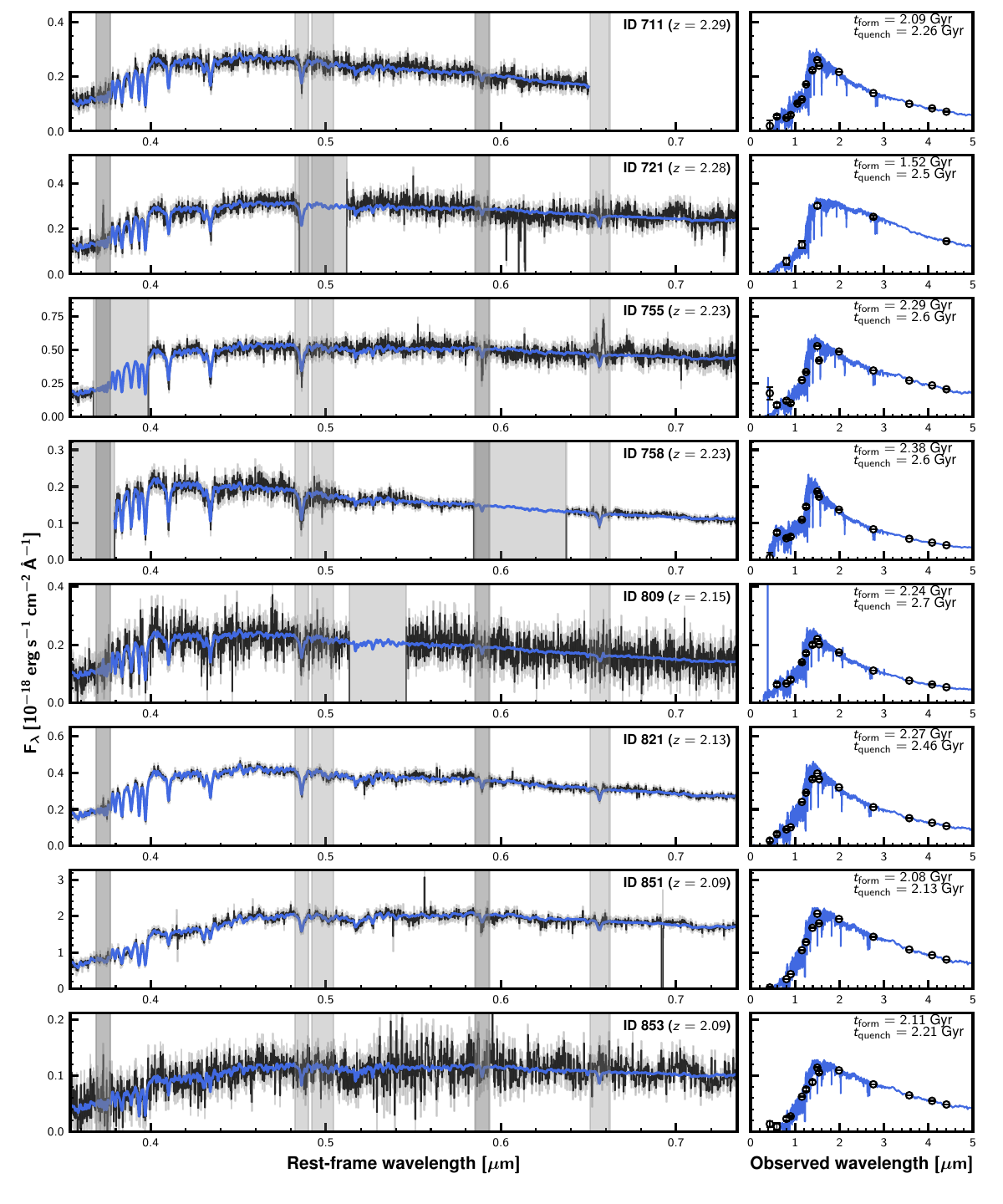}
    \caption{Left panels: JWST/NIRSpec spectra of massive quiescent galaxies used for comparison in Section \ref{subsec:SFHsim}. The black line and gray shaded region correspond to the observed spectrum and its $1\sigma$ uncertainty, respectively. The blue line corresponds to the best-fit {\sc Bagpipes}. The masked regions during the fitting and detector gaps are shown as the gray vertical rectangles. Right panels: Their photometric SEDs. The blue line is the best-fit SED obtained with {\sc Bagpipes}.}
    \label{fig:bagpipes_massiveQG1}
\end{figure*}

\begin{figure*}
    \centering
    \includegraphics[width=0.9\linewidth]{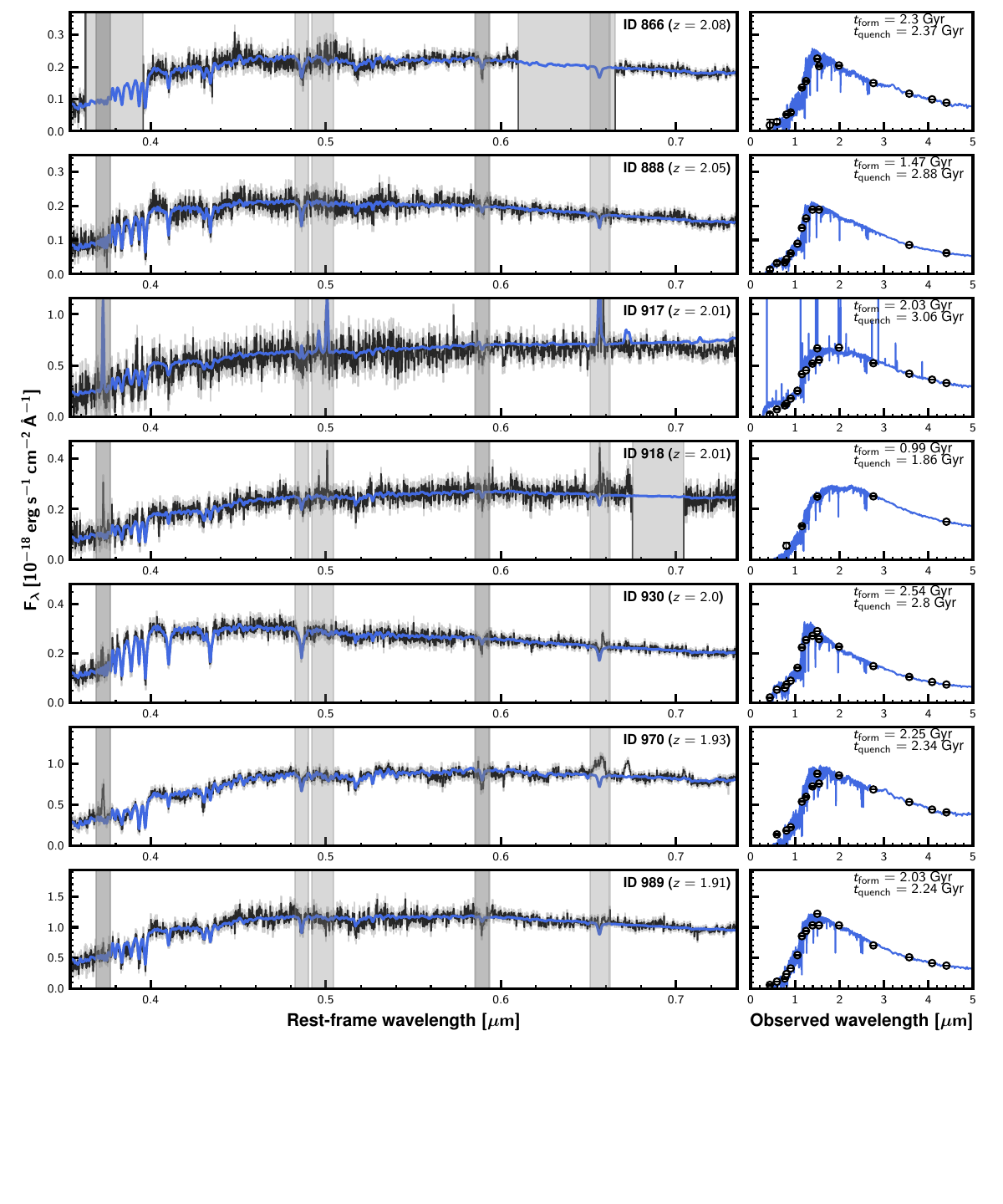}
    \caption{Continued.}
    \label{fig:bagpipes_massiveQG2}
\end{figure*}

\section{Galaxies around \name}\label{app:env}
\begin{figure*}
    \centering
    \includegraphics[width=1\linewidth]{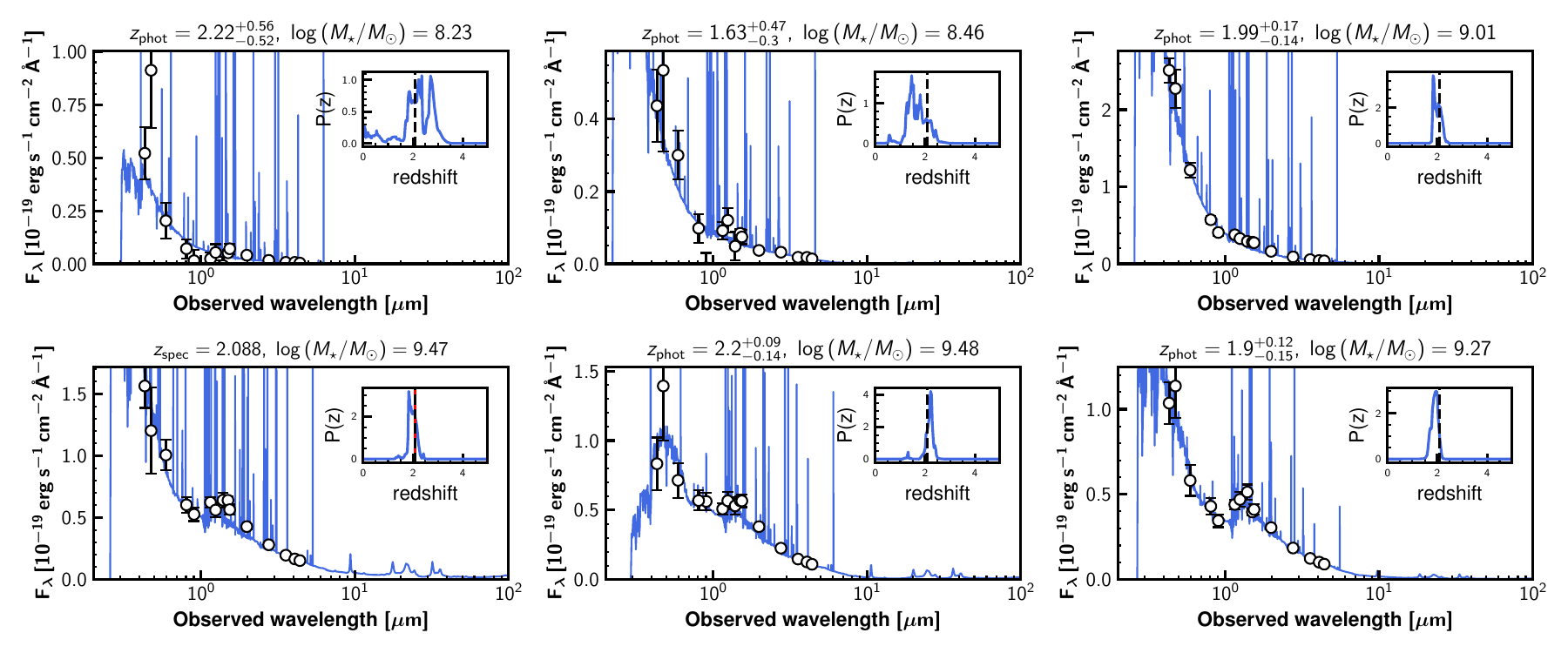}
    \caption{Best-fit SED of six galaxies within 50 pkpc from \name\ and with consistent redshifts. In each panel, the inset shows the probability distribution of their photometric redshift. The black dashed line corresponds to the spectroscopic redshift of \name. One source at lower left is spectroscopically confirmed, and its spectroscopic redshift is shown as a red line.}
    \label{fig:EnvEAZYSED}
\end{figure*}

\begin{figure}
    \centering
    \includegraphics[width=0.8\linewidth]{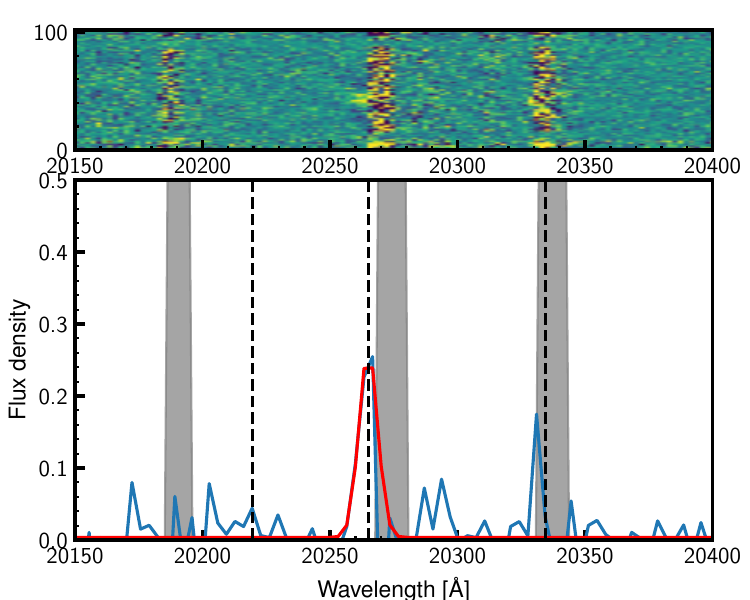}
    \caption{Keck/MOSFIRE K-band spectrum around an emission line in one of the galaxies close to the \name. The top panel shows the 2D spectrum, and the bottom panel shows the 1D extraction and its $1\sigma$ uncertainty in blue. The gray shaded regions mark wavelengths severely affected by sky lines. The best-fit Gaussian function used to determine the spectroscopic redshift is shown in red line.}
    \label{fig:KeckHalpha}
\end{figure}

Figure \ref{fig:EnvEAZYSED} shows the best-fit SED of six galaxies with photometric redshift consistent with that of \name\ and within a 50 pkpc radius (Section \ref{subsec:env}). Figure \ref{fig:KeckHalpha} shows the Keck/MOSFIRE K-band spectrum of one of these galaxies, showing a prominent emission line at $\lambda_{\rm obs}\sim2.027\, {\rm \mu m}$ detected at $6\sigma$ significance. This spectrum was taken as part of the ZFIRE survey \citep{Nanayakkara_2016}. A Gaussian function is fitted to this emission line to derive the line center and total flux. Considering the redshift probability distribution function from {\sc eazy-py} modeling, we conclude that the emission line corresponds to H$\alpha$, fixing the spectroscopic redshift to $z_{\rm spec} = 2.0878\pm0.0002$.

\end{appendix}

\end{document}